\documentclass[aps,twocolumn,showpacs,preprintnumbers,amsmath,amssymb]{revtex4}

\usepackage{graphicx}
\usepackage{dcolumn}
\usepackage{bm}

\newcommand{\ie}{{\it i.e. }}
\newcommand{\eg}{{\it e.g. }}

\newcommand{\vs}{{\it vs. }}

\newcommand{\etal}{{\it et~al. }}

\newcommand{\gapprox}{{\scriptscriptstyle\stackrel{>}{\sim}}}

\newcommand{\fig}[1]{Fig.~\ref{#1}}

\newcommand{\co}[2]{\ifcase #1 \or #2 \fi}
\newcommand{\nbal}{Nb/Al-AlO$_x$/Nb}

\begin{document}


\title{3-junction SQUID rocking ratchet}

\author{A.~Sterck}
\author{R.~Kleiner}%
\author{D.~Koelle}%
 \email{koelle@uni-tuebingen.de}
\affiliation{%
Physikalisches Institut -- Experimentalphysik II, Universit\"{a}t T\"{u}bingen, \\
Auf der Morgenstelle 14, D-72076 T\"{u}bingen, Germany
}%


\date{\today}

\begin{abstract}
We investigate 3-junction SQUIDs which
show voltage rectification if biased with
an ac current drive with zero mean value.
The Josephson phase across the SQUID
experiences an effective ratchet
potential, and the device acts as an
efficient rocking ratchet, as
demonstrated experimentally for adiabatic
and nonadiabatic drive frequencies.
For high-frequency drives the rectified
voltage is quantized due to
synchronization of the phase dynamics
with the external drive.
The experimental data are in excellent
agreement with numerical simulations
including thermal fluctuations.
\end{abstract}

\pacs{05.40.-a, 05.60.-k, 74.40.+k, 85.25.Dq} 

\maketitle


During the last decade, directed
molecular motion in the absence of a
directed net driving force or temperature
gradient in biological systems has drawn
much attention to Brownian
motors.\cite{Reimann02, Linke02}
Nonequilibrium fluctuations can induce
e.g. transport of particles along periodic
structures which lack reflection symmetry
-- so-called ratchets.
An important class of ratchets is given by the {\em
rocking ratchet}, characterized by a time-independent
potential and an external perturbation (driving force)
which may be either deterministic or stochastic, or a
combination of
both.\cite{Magnasco93,Bartussek94,Faucheux95,
Mateos00,Borromeo02}
The number of ratchet systems considered
for experimental studies has steadily been
growing during recent years.\cite{Linke02}
In particular, superconducting ratchets, based on the
motion of Abrikosov vortices
\cite{Wambaugh99,Lee99a,Olson01,
Villegas03,VandeVondel05}, Josephson
vortices\cite{Trias00,Goldobin01,
Carapella01a,Carapella02b,Beck05unpub} or the phase
difference of the superconducting wavefunction
(Josephson phase) in SQUID
ratchets\cite{Zapata96,Weiss00,Sterck02, Berger04}
have been investigated.
Those systems offer the advantage of (i) good
experimental control over externally applied driving
forces (here: currents), (ii) easy detection of
directed motion, which creates a dc voltage, and (iii)
experimental access to studies over a wide frequency
range of external perturbations (adiabatic and
non-adiabatic regime), and transition from overdamped
to underdamped dynamic regimes, enabling studies of
inertial effects and transition to chaos.

Zapata \etal\cite{Zapata96} proposed a
3-junction (3JJ) SQUID ratchet, which
consists of a superconducting loop of
inductance $L$, intersected by one
Josephson junction in one arm and by two
Josephson junctions connected in series
in the other arm.
For vanishing $L$ a quasi-one dimensional
(1D) ratchet potential can be obtained,
and rectification of an ac bias current
for low- and high-frequency drive of such
a rocking ratchet has been predicted.
In this paper we investigate a 3JJ SQUID, similar to
the one proposed in\cite{Zapata96}. We derive the
equations of motion and the ratchet potential for our
type of device, we present its experimental
realization, and we investigate its operation as a
rocking ratchet for both adiabatic and non-adiabatic
drive.
We also compare experimental results with
numerical simulations for our device and
for the originally proposed 3JJ SQUID
ratchet.


\begin{figure}[tb]
\center{
\includegraphics[width=\columnwidth,clip]
{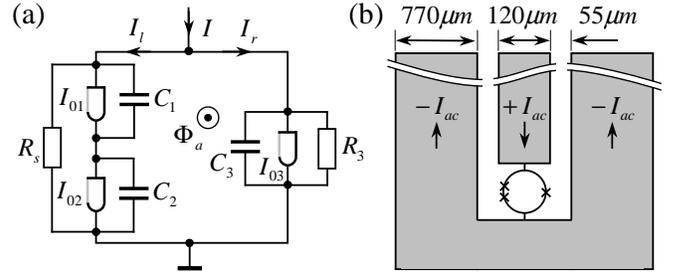}} \caption{Three-junction SQUID: Equivalent
circuit (a) and schematic layout (b).
\label{fig:networks}}
\end{figure}
We first discuss the underlying dynamic
equations, using the resistively and
capacitively shunted junction
model\cite{Stewart68,Mccumber68}.
To simplify the layout, junctions 1,2 in the left arm
are shunted by a {\it common} resistor $R_s$, in
contrast to the original proposal\cite{Zapata96},
where junctions 1,2 are shunted individually.
With Kirchhoff's laws, the Josephson
equations and the phase differences
$\delta_k$ across junction $k=1,2$, the
current $I_l$ through the left arm is
\begin{subequations}
\begin{equation}
\label{eq:motion-a}
\frac{I}{2}+J
= I_l
= \frac{\Phi_0 C_k}{2\pi}\ddot{\delta_k}
 +\frac{\Phi_0}{2\pi R_s}
  (\dot{\delta_1}+\dot{\delta_2})
 +I_{0k} \sin{\delta_k}+I_{Ns}.
\end{equation}
$\Phi_0=h/2e$ is the flux quantum; $I_{Ns}$
is the Nyquist noise current (with
spectral density $S_I(f)=4k_BT/R_s$)
produced by the shunt.
We neglect the (subgap) resistances of the
unshunted junctions which are much larger
than $R_s$.
Note that the original model\cite{Zapata96}
can be obtained from (\ref{eq:motion-a})
by replacing
$\Phi_0(\dot{\delta_1}+\dot{\delta_2})/2\pi R_s$
by
$\Phi_0\dot{\delta_k}/2\pi R_k$.
The current $I_r$ through the right arm is
\begin{equation}
\label{eq:motion-b}
\frac{I}{2}-J
= I_r
= \frac{\Phi_0 C_3}{2\pi}\ddot{\delta_3}
 +\frac{\Phi_0}{2\pi R_3}\dot{\delta_3}
 +I_{03} \sin{\delta_3}+I_{N3}\;.
\end{equation}
\end{subequations}
The total bias current is $I=I_l+I_r$, and
the Langevin equations(\ref{eq:motion-a}),
(\ref{eq:motion-b}) are coupled via the
circulating current around the loop
$J=(I_l-I_r)/2$.
The phase differences $\delta_k$ are
connected via
\begin{equation}
\label{eq:motion-c}
\delta_3-(\delta_1+\delta_2)
= \frac{2\pi}{\Phi_0}(\Phi_a+LJ)
= \frac{2\pi}{\Phi_0}\Phi_T\;.
\end{equation}
The total flux $\Phi_T$ through the SQUID
loop has contributions from the applied
flux $\Phi_a$ and from the circulating
current $J$.
For comparison with experimental results,
we performed numerical simulations to
solve the coupled Langevin equations
(\ref{eq:motion-a}), (\ref{eq:motion-b})
and (\ref{eq:motion-c}).
From the second Josephson relation we
obtain the momentary voltage across the
junctions
($U_k=\Phi_0\dot{\delta_k}/2\pi$) and,
by time averaging, the dc voltage $V$
across the SQUID and thus the current
voltage characteristic (IVC) and the
critical current $I_c$ \vs applied
flux $\Phi_a$.

\begin{figure}[tb]
\center{
\includegraphics[width=\columnwidth,clip]
{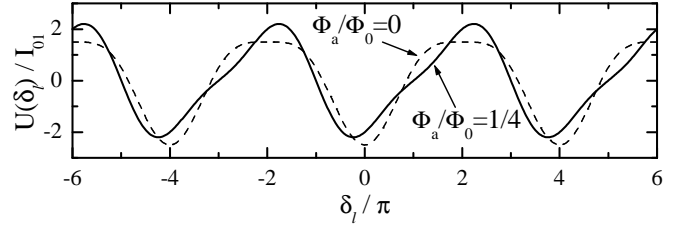}} \caption{Potential $U$ from
eq.(\ref{eq:1d-potential-sym}) for $s=1/2$, and two
values of applied flux $\Phi_a$.}
\label{fig:potential}
\end{figure}

In the following we derive the ratchet
potential in the overdamped limit, \ie
with 
$\beta_{Ck}\equiv 2\pi I_{0k}R_k^2C_{k}\ll 1$
for all three junctions.
The displacement currents
$\propto C_k\ddot{\delta_k}$ in
(\ref{eq:motion-a}), (\ref{eq:motion-b})
can then be neglected and the equations of
motion reduce to only two differential
equations
\begin{subequations}
\begin{align}
\label{eq:motion-left}
I_l&
= \frac{\Phi_0}{2\pi R_s}
  (\dot{\delta_1}+\dot{\delta_2})
 +I_{01}\sin{\delta_1}
 +I_{Ns}\;,\\
\label{eq:motion-right}
I_r&
= \frac{\Phi_0}{2\pi R_3}\dot{\delta_3}
 +I_{03}\sin{\delta_3}
 +I_{N3}\;,
\end{align}
\end{subequations}
plus conservation of the supercurrent in the
left arm
$I_{01}\sin{\delta_1}=I_{02}\sin{\delta_2}$, 
which can be used to express the overall
phase difference for the left arm
$\delta_l \equiv \delta_1+\delta_2$
in terms of $\delta_1$ only
\begin{equation}
\label{eq:deltal}
\delta_l
= \delta_1+\arcsin{(q \sin{\delta_1})}\;,
\end{equation}
with $q \equiv I_{01}/I_{02}$.
Without loss of generality we assume
$I_{01}\le I_{02}$, \ie $q\le 1$.
Equation (\ref{eq:deltal}) can be
reversed to
\begin{equation}
\label{eq:delta1}
\delta_1
= \arctan{
  \Bigl(
  \frac{\sin{\delta_l}}{\cos{\delta_l+q}}
  \Bigr)}\;.
\end{equation}
Inserting (\ref{eq:delta1}) and the
expression for the circulating current
(\ref{eq:motion-c}) into
(\ref{eq:motion-right}) and
(\ref{eq:motion-left}) leads to
\begin{subequations}
\begin{eqnarray}
\label{eq:total}
\frac{\Phi_0}{2\pi R_3}\dot{\delta_3}
  -\frac{I}{2}
  +I_{N3}
&=&
-I_{03}\sin{\delta_3}
  -\frac{\Phi_0}{2\pi L}
  (\delta_3-\delta_l
   -2\pi\frac{\Phi_a}{\Phi_0})
  \notag\\
&\equiv&
  -\frac{\partial U}{\partial\delta_3},\\
\frac{\Phi_0}{2\pi R_s}\dot{\delta_l}
  -\frac{I}{2}
  +I_{Ns}
&=&
-I_{01}\sin{(\arctan{
   \Bigl(\frac{\sin{\delta_l}}{\cos{\delta_l}+q}
   \Bigr)})}\\
&&
 +\frac{\Phi_0}{2\pi L}
  (\delta_3-\delta_l-2\pi\frac{\Phi_a}{\Phi_0})
\equiv -\frac{\partial U}{\partial\delta_l}\;.\notag
\end{eqnarray}
\end{subequations}
The potential $U(\delta_l,\delta_3)$ 
is obtained by integration as
\begin{eqnarray}
\label{eq:2d-potential}
U(\delta_l,\delta_3)
&=&
 -s I_{01}\cos{\delta_3}
 +\frac{\Phi_0}{4\pi L}(\delta_3-\delta_l-2\pi\frac{\Phi_a}{\Phi_0})^{2}\notag\\
&&
 \mp\frac{I_{01}}{q}(\sqrt{1+2q\cos{\delta_l}+q^{2}}+q-1)\;.
\end{eqnarray}
The minus (plus) sign in the second line
applies for $\cos(\delta_l/2)>0$ ($<0$).
The symmetry of $U$ is determined by the
applied flux and $s\equiv I_{03}/I_{01}$.
If the screening parameter
$\beta_{L}\equiv 2LI_{0}/\Phi_0\ll 1$
(with $I_0\equiv(I_{03}+I_{01})/2$), the
phase differences are fixed by the applied
flux $\Phi_a$ \ie
$\delta_3-\delta_l=2\pi\Phi_a/\Phi_0$;
hence $\delta_3$ can be eliminated:
\begin{eqnarray}
\label{eq:1d-potential}
U(\delta_l)
&=&
  -s I_{01}
   \cos{(\delta_l+2\pi\frac{\Phi_a}{\Phi_0})}
   \notag\\
&&
  \mp\frac{I_{01}}{q}
  (\sqrt{1+2q\cos{\delta_l}+q^{2}}+q-1)\;.
\end{eqnarray}
If the two junctions in the left arm have
equal critical currents ($q=1$) 
the above expression turns into
\begin{equation}
\label{eq:1d-potential-sym}
U(\delta_l)
= -sI_{01}\cos{(\delta_l+2\pi\frac{\Phi_a}{\Phi_0})}
  -2I_{01}\cos{(\delta_l/2)}\;,
\end{equation}
which coincides with the result presented
in\cite{Zapata96} and is shown in
\fig{fig:potential} for $s=1/2$ (the value
proposed in\cite{Zapata96}).


As one can see from \fig{fig:potential},
it is the applied flux that breaks the
reflection symmetry of the potential.
The critical currents $I_c^+(\Phi_a)$,
$I_c^-(\Phi_a)$ represent the force to
overcome the potential barrier for
positive and negative bias current
polarity, respectively.
Hence, measuring $I_c^\pm$ provides a way
to probe the asymmetry of the potential.
The maximum of $I_c(\Phi_a)=2I_0$ is
reached if each arm carries a current,
that is equal to its critical current,
\ie $I_r=I_{03}$ and $I_l=I_{01}$.
The circulating current is
then given by $J=(I_{01}-I_{03})/2$, and
the three phase differences have the
values $\delta_1=\delta_3=\pi/2$ and
$\delta_2=\arcsin q \le\pi/2$.
Inserting this into (\ref{eq:motion-c})
and solving for $\Phi_a$ gives the
applied flux $\Phi_a^\pm$ at maximum
$I_c$ ($+$ and $-$ corresponds to
positive and negative bias current,
respectively)
\begin{equation}
\label{eq:max-ic}
\Phi_a^\pm
=
\mp\frac{\Phi_0}{4}
   \bigl(\frac{2}{\pi}\arcsin q
         + 2\beta_L\frac{1-s}{1+s}\bigr)\;.
\end{equation}
If $q\approx 1$, the first term yields a
shift $\approx\mp\Phi_0/4$ of the
$I_c^\pm(\Phi_a)$-dependence.
The second term has the same sign and is
proportional to $\beta_L$.
Equation (\ref{eq:max-ic}) can be used to
estimate $q$, $\beta_L$ and $s$ from the
measured $I_c(\Phi_a$).


The 3JJ SQUIDs we investigated were
fabricated at HYPRES\footnote{Hypres,
Elmsford (NY), USA.
http://www.hypres.com} using \nbal\,
junctions of nominal critical current
density $j_0=1\,\rm{kA/cm^2}$ at
$T$=4.2\,K, capacitance per area
$C'=38\,\rm{fF/\mu m^2}$ and nominally
identical shunt resistors $R_s=R_3$.
Junction areas $A_{J,1}=A_{J,2}
=3.8\times3.8\mu\rm{m}^2$, and
$A_{J,3}=2.6\times2.6\mu\rm{m}^2$,
correspond to $s\approx 1/2$ and $q=1$.
For microwave irradiation (up to
$28\,\rm{GHz}$), the SQUIDs are
integrated in a coplanar waveguide (see
\fig{fig:networks}(b)) with $50\,\Omega$
impedance.
In total we investigated six devices
which differed by the size of the SQUID
hole, {\em i.e.} by the SQUID inductance
$L$.
All devices showed very similar behavior.
Below we discuss only one device with a
$2\times3\mu\rm{m^2}$ hole.
All measurements were performed at
$T=4.2\,K$ in a magnetically and
electrically shielded environment, with
low pass filters in the voltage leads.
Apart from microwave drives we also
studied the ratchet behavior in the kHz
regime with ac currents directly fed into
the bias leads.

\begin{figure}[tb]
\center{
\includegraphics[width=\columnwidth,clip]
{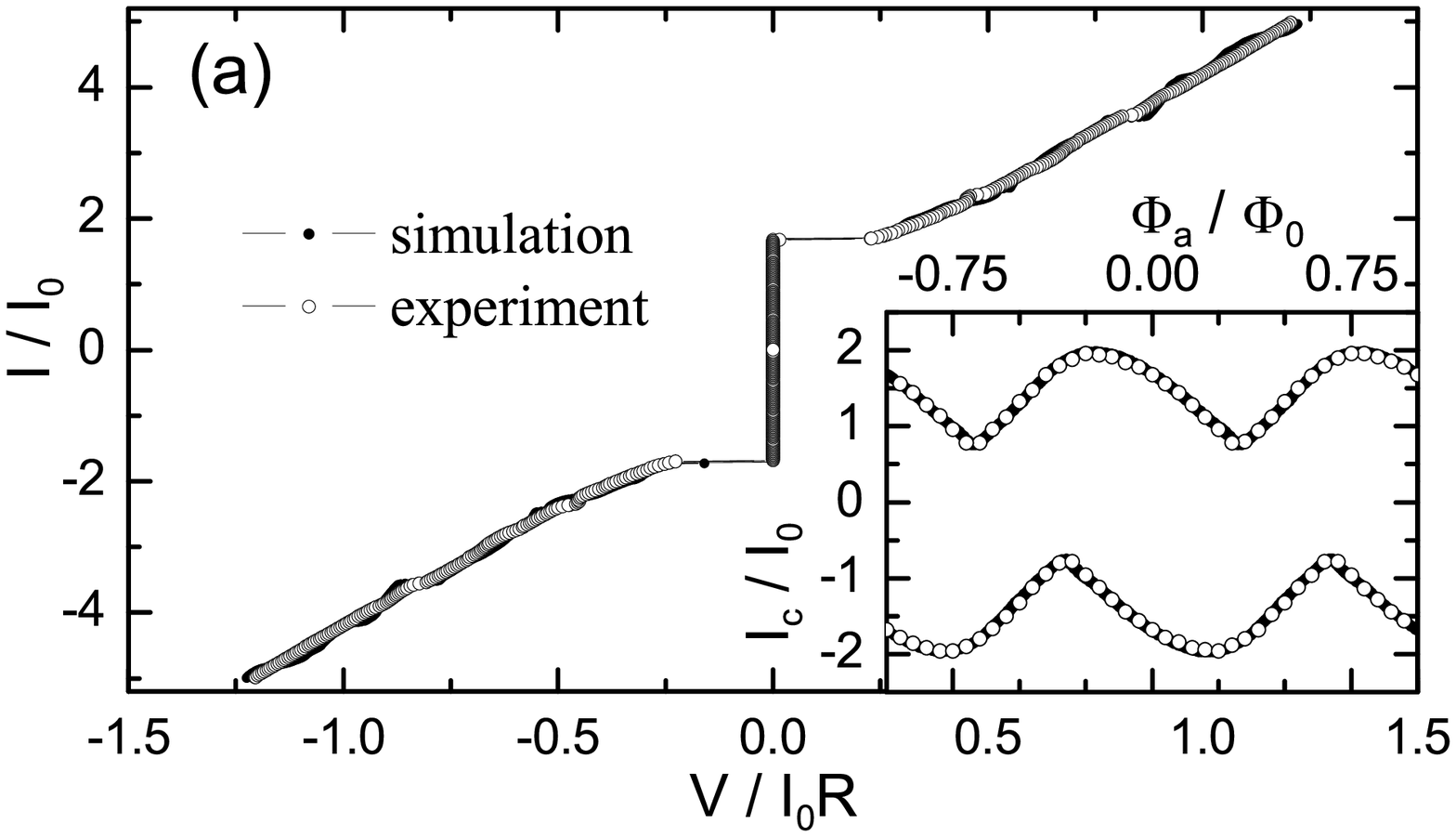}
\includegraphics[width=\columnwidth,clip]
{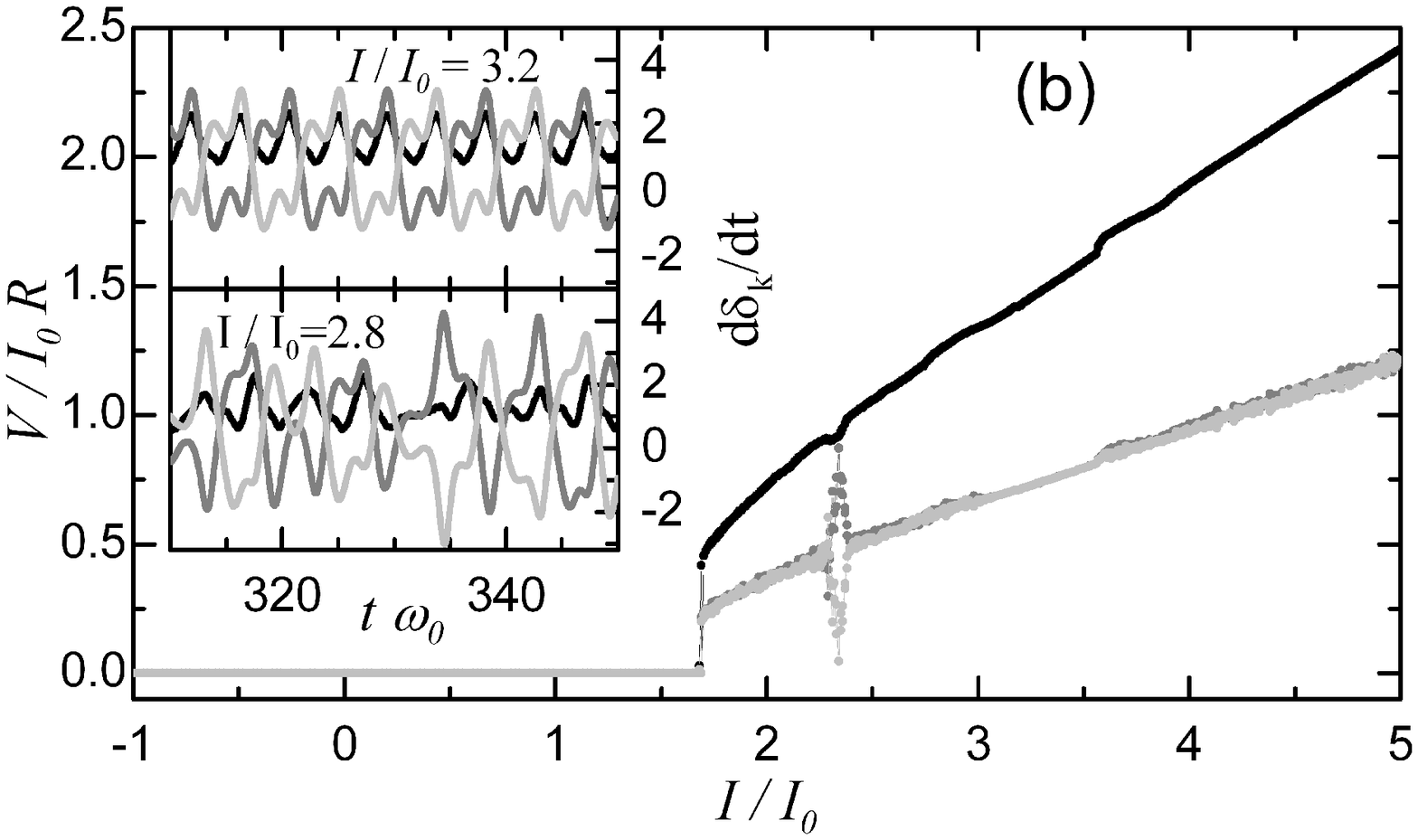}} \caption{SQUID characteristics without ac
drive:
(a) $I(V)$-curves ($\Phi_a=0$) and
$I_c(\Phi_a)$ (inset).
(b) Simulated $V(I)$-curve (black) and
individual voltages across junction 1
(grey) and 2 (light grey).
Insets: Voltage vs. time for junction 3
(black), 1 (grey) and 2 (light grey) for
$I/I_0 = 3.2$ and $2.8$.
Parameters for numerical simulations:
$\Gamma=2\cdot 10^{-3}$,
$\beta_{C,1}=\beta_{C,2}=0.38$,
$\beta_{C,3}=0.09$, $\beta_L=0.1$,
$s=0.5$ and $q=0.99$.}
\label{fig:dc-properties}
\end{figure}

\fig{fig:dc-properties}(a) shows a
measured IVC ($\Phi_a=0$) with normal
resistance $R_n\approx 0.7\,\Omega$, and
thus $R_s\approx R_3\equiv R
\approx 1.4\,\Omega$.
The inset shows 
the critical currents $I_c^\pm(\Phi_a)$ 
with maximum $I_c=2I_0=186\mu\rm{A}$,
giving a noise parameter
$\Gamma\equiv 2\pi
k_BT/(I_0\Phi_0)\approx 2\cdot 10^{-3}$,
characteristic voltage $V_c\equiv
I_0R=130\,\mu\rm V$ and characteristic
frequency $f_c\equiv V_c/\Phi_0=63\,\rm
GHz$.
With the design values for the
junction capacitances $C_1=C_2=0.54\,$pF
and $C_3=0.27\,$pF we obtain
$\beta_{C,1}=\beta_{C,2}=0.38$ and
$\beta_{C,3}=0.09$.
The shape (asymmetry) and modulation
depth of $I_c(\Phi_a)$ is determined by
$s$, $q$ and $\beta_L$.
Adjusting these parameters in numerical
simulations of $I_c(\Phi_a)$ (with values
for $\Gamma$ and $\beta_{Ck}$ as
determined above), we find best agreement
with the measured $I_c(\Phi_a)$ 
for $\beta_L=0.1$, $s=0.5$ and $q=0.99$.
These parameters are very close to the
design values, and have been used later to
numerically calculate the response of the
SQUID to an ac bias current.

As a prominent feature in the experimental
zero field IVCs, we observe step-like
structures in the resistive state;
simulated IVCs exhibit the same generic
shape [cf.~\fig{fig:dc-properties}(a)].
The simulation allows to trace
the individual voltage drops across the
junctions, as shown in
\fig{fig:dc-properties}(b).
It turns out that the step structures in
the IVCs can be associated with switching
of junctions 1 and 2 between different
states:
(i) one junction is in the zero voltage
state while the other carries a large
voltage;
(ii) both junctions are in the voltage
state and oscillate coherently, \ie with
the same frequency;
typically these oscillations are out of
phase [cf.~upper inset in
\fig{fig:dc-properties}(b)];
(iii) both junctions are in the voltage
state;
however, oscillations are irregular and
incoherent [cf.~lower inset in
\fig{fig:dc-properties}(b)].
This state is associated with a more ''noisy'' IVC as
compared to (ii).
Our simulations show that over a wide range
of bias currents the junctions oscillate
incoherently and thus are far from the scenario
discussed in\cite{Zapata96}, where junctions 1
and 2 were assumed to behave identical, i.e.
$\delta_1=\delta_2$ at all times
(''synchronous state'').

We performed extensive numerical
simulations to compare our device (with
common shunt for junctions $1,2$) with
the one in \cite{Zapata96}.
For $T=0$ both devices exhibit phase synchronous
oscillations over the entire range of bias currents if
junctions 1,2 are identical and with initial values
$\delta_k=0$ and $\dot{\delta}_k=0$.
However, choosing either different initial
conditions or non-identical junction
parameters results in a behavior very
similar to the one shown in
\fig{fig:dc-properties}(b).
At $T=0$ the synchronous state is more
easily destroyed if junctions 1,2 have a
common shunt, as compared to being
shunted individually.
In the latter case, the synchronous state is more
stable, due to a damping term
$\propto\partial(\delta_1-\delta_2)/\partial t$ which
is absent in the equations of motion for the system
with common shunt for junctions
1,2\cite{Sterck05unpub}.
However, thermal fluctuations at finite
temperature, also tend to destroy the
synchronous state for the individually
shunted 3JJ SQUID.
Hence, for the experimentally relevant case, {\it
i.e.} at finite temperature ($\Gamma\gapprox 10^{-3}$)
and slight junction asymmetries ($q\neq 0$), both
types of SQUID ratchets show typically incoherent
oscillations of junctions 1,2.

\begin{figure}[tbh]
\center{
\includegraphics[width=\columnwidth]
{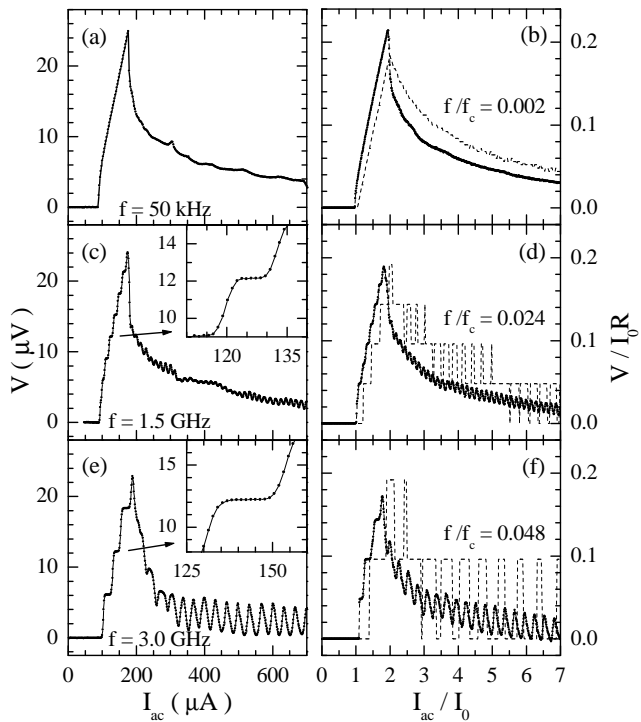}} \caption{Rectified voltage \vs ac drive
amplitude at $\Phi_a=\Phi_0/4$ for a 3JJ SQUID ratchet
with increasing drive frequency $f$ (from top to
bottom).
Left column shows experimental data at
$T=4.2\,\rm{K}$ (insets in (c),(e) show
enlarged section of one voltage step) and
right column shows the corresponding
simulated curves (parameters as in
\fig{fig:dc-properties}).
The dashed lines in (b), (d), (f) were
calculated for $T=0$, with $I_{01}=I_{02}$ ($q=1$) and
$\delta_k=\dot{\delta}_k=0$ ($k=1,2$) at $t=0$ .
\label{fig:ratchet}}
\end{figure}

\fig{fig:ratchet} shows experimental
data, and numerical simulation results,
for the rectified voltage $V$ under ac
drive $I=I_{ac}\sin2\pi f t$, clearly
demonstrating operation of the SQUID as a
rocking ratchet.
For adiabatic drive $f\ll f_c$
[c.f.~\fig{fig:ratchet}(a)] rectification
appears at  $I_{ac}\gapprox I_0$.
$V$ sharply peaks at $I_{ac}\approx 2I_0$
and decreases again with increasing
$I_{ac}$, as predicted for a 1D rocking
ratchet\cite{Bartussek94} and as observed
experimentally for asymmetric (two
junction) dc SQUID
ratchets\cite{Weiss00,Sterck02}.
In contrast to the prediction for
overdamped rocking ratchets
\cite{Bartussek94,Borromeo02}, the
initial increase in $V(I_{ac})$ up to the
maximum $V_{max}$ is not linear. Instead,
it sets off with a very steep slope
$dV/dI_{ac}$ which becomes smaller at
higher voltage.
A strikingly similar behavior has been
predicted for a strongly underdamped
rocking ratchet \cite{Borromeo02}.
In our case, the motion of $\delta_l$ is
overdamped; however for
$\delta_1-\delta_2$ it is underdamped,
which obviously gives the same result as
reported in \cite{Borromeo02}.
$V_{\rm max}=25\mu\rm{V}$ is as large as
20\% of $V_c$, in excellent agreement
with simulations shown in
\fig{fig:ratchet}(b).
As maximum rectification requires
$\Phi_a^+-\Phi_a^-\approx\Phi_0/2$, it is
easy to understand from
eq.(\ref{eq:max-ic}) why the 3JJ SQUID
ratchet is superior to the asymmetric dc
SQUID ratchet:
For the asymmetric dc SQUID
the arcsin-term in (\ref{eq:max-ic})
is absent, and the shift in the
$I_c^\pm(\Phi_a)$-curves is determined
by $\beta_L$ and $s$ only.
Hence, the condition for maximum rectification
depends strongly on the SQUID parameters
and is more difficult to fulfill experimentally.
In contrast, for the 3JJ SQUID, it is
sufficient to keep $\beta_L$ small, so
that the arcsin-term dominates.
For $q\approx 1$, the condition for maximum
rectification is then easy to achieve
experimentally.

At nonadiabatic drive frequencies 
[c.f.~\fig{fig:ratchet}(c),(e)] we observe
a step-like increase of $V(I_{ac})$ up to
$V_{max}$ and oscillations in $V(I_{ac})$
for higher drive amplitudes.
The voltage steps appear at
$V/(I_0R)=n\cdot(f/f_c)$ ($n$: integer)
and can be interpreted as Shapiro steps,
where the phase dynamics synchronizes with
the external drive\cite{Sterck02}.
This {\em quantization} of the ratchet
effect has been predicted as a
characteristic feature of
nonadiabatically driven 1D rocking
ratchets \cite{Bartussek94,Zapata96};
however, it has not been observed
experimentally until now, due to thermal
noise smearing\cite{Sterck02}.
For the same reason, $V(I_{ac})$ oscillates at
higher drive amplitudes, rather than showing
a step-like behavior.
This effect of thermal noise was also
predicted in\cite{Bartussek94} for 1D
rocking ratchets and is clearly
demonstrated for our device in the
simulations of $V(I_{ac})$ shown in
\fig{fig:ratchet}(d), (f) for finite and
zero $T$.
For finite $T$, our
simulations nicely reproduce
the experimental data.
For the case $T=0$, simulations were
performed with $q=1$ and identical
initial conditions
$\delta_{1,2}=\dot{\delta}_{1,2}=0$.
This leads to the formation of the
synchronous state (\ie
$\delta_1=\delta_2$ at all times).
Only in this case a step-like switching
of the rectification is observed in the
numerical simulations also at large ac
drive amplitudes; \ie experimental
observation of quantized rectification
for large driving amplitudes, will be
quite difficult.
Furthermore, we note that only in the synchronous
state ($\varphi\equiv\delta_1-\delta_2=0$) the
dynamics of the system is described by $\delta_l(t)$
only, as the motion of a particle in the 1-dimensional
potential (\ref{eq:1d-potential-sym}) which is
$4\pi$-periodic in $\delta_l$. In contrast, if
$\dot{\varphi}\neq 0$, the dynamics can be described
in a 2-dimensional potential along $\delta_l$ and
$\varphi$ \cite{Sterck05unpub}. This potential has
minima which are separated by $2\pi$ if projected onto
the $\delta_l$-axis. This explains the doubling of the
Shapiro step height calculated for the synchronous
state [c.f.~dashed lines in
Figs.~\ref{fig:ratchet}(d),(f)] as compared to the
calculations for finite $T$ and as observed
experimentally [c.f.~Figs.\ref{fig:ratchet}(c),(e)].



In conclusion, following a proposal
in\cite{Zapata96}, we have realized 3JJ
SQUIDs and demonstrated operation of
those devices as very efficient rocking
ratchets.
In contrast to the original proposal,
which assumed that the
two junctions connected in series
oscillate in-phase, we found that these
junctions usually oscillate incoherently.
Nonetheless the devices show a large
ratchet effect, more than two orders of magnitude
above initial estimates given in\cite{Zapata96}.
Apart from rectification of a harmonic low-frequency
drive, we find rectification of nonadiabatic drives
with striking properties, such as Shapiro-like steps,
\ie quantization of the velocity of directed motion,
which depends only on the drive frequency.
The simplicity of the design, the good experimental
control over external drives and device parameters
which define \eg the shape of the ratchet potential,
and, finally, easy detection of directed motion by
measuring voltage, gives this device excellent
perspectives for further basic experimental studies
of ratchet effects, \eg for stochastic drives.

We gratefully acknowledge financial
support from the Deutsche
Forschungsgemeinschaft.

\bibliography{ratch,jj,kk,koelle-only,unpub}

\end{document}